\documentclass[12pt]{article}

\def\bra#1{{\langle#1|}}
\def\ket#1{{|#1\rangle}}
\def\CN{{\cal N}}
\def\CR{{\cal R}}
\def\CI{{\cal I}}
\def\CT{{\cal T}}
\def\CE{{\cal E}}
\def\CD{{\cal D}}
\def\CW{{\cal W}}
\def\CV{{\cal V}}
\def\CK{{\cal K}}
\usepackage{graphicx}
\usepackage{scicite}
\usepackage{amsmath}

\usepackage{times}

\newenvironment{sciabstract}{%
\begin{quote} \bf}
{\end{quote}}

\newcounter{lastnote}
\newenvironment{scilastnote}{%
\setcounter{lastnote}{\value{enumiv}}%
\addtocounter{lastnote}{+1}%
\begin{list}%
{\arabic{lastnote}.}
{\setlength{\leftmargin}{.22in}}
{\setlength{\labelsep}{.5em}}}
{\end{list}}

\title{Quantum Steganography}

\author
{Bilal A. Shaw,$^{1\ast}$ Todd A. Brun$^{2}$\\
\\
\normalsize{$^{1}$Computer Science Department, University of Southern California,}\\
\normalsize{$^{2}$Electrical Engineering Department, University of Southern California}\\
\normalsize{Los Angeles, CA 90089, USA}\\
\\
\normalsize{$^\ast$To whom correspondence should be addressed; E-mail:  bilalshaw@gmail.com.}
}

\date{}

\begin{document} 


\baselineskip24pt


\maketitle


\begin{sciabstract}
Steganography is the process of hiding secret information by embedding it in an ``innocent'' message.  We present protocols for hiding quantum information in a codeword of a quantum error-correcting code passing through a channel.  Using either a shared classical secret key or shared entanglement the sender (Alice) disguises her information as errors in the channel.  The receiver (Bob) can retrieve the hidden information, but an eavesdropper (Eve) with the power to monitor the channel, but without the secret key, cannot distinguish the message from channel noise.  We analyze how difficult it is for Eve to detect the presence of secret messages, and estimate rates of steganographic communication and secret key consumption for certain protocols.
\end{sciabstract}



Steganography is the science of hiding a message within a larger innocent-looking plain-text message, and communicating the resulting data over a communications channel or by a courier so that the steganographic message is readable only by the intended receiver.  The word comes from the Greek words \textit{steganos} which means ``covered,'' and \textit{graphia} which means ``writing.''  The art of information hiding dates back to 440 B.C. to the Greeks \cite{Herodotus}.  The term steganography was first used in 1499 by Johannes Trithemius in his \textit{Steganographia}, which was one of the first treatises on the use of cryptographic and steganographic techniques~\cite{Trithemius}.

The modern study of steganography was initiated by Simmons and the paradigm can be stated as follows~\cite{Simmons}.  Alice and Bob are imprisoned in two different cells that are far apart.  They would like to devise an escape plan, but the only way they can communicate with each other is through a courier who is under the command of the warden (Eve, the adversary) of the penitentiary.  The courier leaks all information to the warden.  If the warden suspects that either Alice or Bob are conspiring to escape from the penitentiary, she will cut off all communication between them, and move both of them to a maximum security cell.  Prior to their incarceration Alice and Bob had access to a shared secret key---assumed to be a sufficiently long string of random bits---which they later exploit to send secret messages hidden in a cover text.  Can Alice and Bob devise an escape plan without arousing the suspicion of the warden?  

Julio Gea-Banacloche~\cite{Banacloche} introduced the idea of hiding secret messages in the form of error syndromes by deliberately applying correctable errors to a quantum state encoded in the three-bit repetition quantum error-correcting code (QECC).  In his paper, however, he did not address the issue of an innocent-looking message---in the protocol he proposed, the messages would not resemble a plausible quantum channel.  The latter is one of the major contributions of our work.  Curty et. al. propose three different quantum steganographic protocols~\cite{Curty}.  However, none of these protocols address the issue of communicating an innocent message over a noisy classical channel or a general quantum channel, or give key-consumption rates.  Natori provides a rudimentary treatment of quantum steganography which is a modification of super-dense coding~\cite{Natori}.  Martin introduced a notion of quantum steganographic communication in~\cite{Martin}.  His protocol is a variation of Bennett and Brassard's quantum-key distribution protocol (QKD), in which he hides a steganographic channel in the QKD protocol.

Our treatment of quantum steganography is more general than those above.  We provide a protocol (Fig. 1) for hiding quantum information using typical sequences of errors for general quantum channels.  We begin by showing how quantum information can be hidden in the noise of a depolarizing channel, using a shared classical secret key between Alice and Bob.  In our first quantum steganographic protocol the channel is intrinsically noiseless (i.e., all noise is controlled by Alice), and in the second case the channel has its own intrinsic noise (not controlled by Alice and Bob).  We calculate the amount of secret key consumed.  We later present a quantum steganographic protocol for general quantum channels.  We also discuss whether Alice and Bob can send a finite amount of hidden information, or can actually communicate at a nonzero asymptotic rate (given an arbitrarily large secret key).  This depends on Eve's knowledge of the physical channel, and Alice and Bob's knowledge of Eve's expectations.  Finally, we address the question of security.  This is two-fold: first, can Eve detect that a secret message has been sent?  And second, can she read the message?

The quantum analog of the classical \textit{binary symmetric channel} (BSC) is the {\it depolarizing channel} (DC) which is one of the most widely used quantum channel models: 
\begin{equation}
\rho \rightarrow \CN\rho = (1-p) \rho + \frac{p}{3} X\rho X + \frac{p}{3} Y\rho Y + \frac{p}{3} Z\rho Z~.
\label{DC}
\end{equation}
That is, each qubit has an equal probability of undergoing an $X$, $Y$, or $Z$ error.  Applying this channel repeatedly to a qubit will map it eventually to the maximally mixed state $I/2$.  We can rewrite this channel in a different but equivalent form:
\begin{equation}
\CN = (1-4p/3) \CI + (4p/3) \CT~.
\label{twirl}
\end{equation}
where  $\CI\rho = \rho$ and $\CT\rho = (1/4)(\rho + X\rho X + Y\rho Y + Z\rho Z)$.  The operation $\CT$ is {\it twirling}:  it takes a qubit in any state $\rho$ to the maximally mixed state $I/2$.  If we rewrite the channel in this way, instead of applying $X$, $Y$, or $Z$ errors with probability $p/3$, we can think of removing the qubit with probability $4p/3$, and replacing it with a maximally mixed state.  This picture makes the steganographic protocol more transparent.  We will first assume that the actual physical channel between Alice and Bob is noiseless.  All the noise that Eve sees is due to deliberate errors that Alice applies to her codewords.   
\begin{enumerate}
\item Alice encodes a covertext of $k_c$ qubits into $N$ qubits with an $[[N,k_c]]$ quantum error-correcting code (QECC).
\item From (\ref{twirl}), the DC would maximally mix $Q$ qubits with probability $p_Q$ where
\begin{equation}
p_Q = {N \choose Q} (4p/3)^Q (1-4p/3)^{N-Q}~.
\label{binomial2}
\end{equation}
For large $N$, Alice can send $M = (4/3)pN(1-\delta)$ stego qubits, where $1 \gg \delta\gg \sqrt{(1-4p/3)/(4p/3)N}$.  (The chance of fewer than $M$ errors is negligibly small.)
\item Using the shared random key (or shared ebits), Alice chooses a random subset of $M$ qubits out of the $N$, and swaps her $M$ stego qubits for those qubits of the codeword.  She also replaces a random number $m$ of qubits outside this subset with maximally mixed qubits, so that the total $Q=M+m$ matches the binomial distribution (\ref{binomial2}) to high accuracy.
\item Alice ``twirls'' her $M$ stego qubits using $2M$ bits of secret key or $2M$ shared ebits.  To each qubit she applies one of $I$, $X$, $Y$, or $Z$ chosen at random, so $\rho \rightarrow \CT\rho$.  To Eve, who does not have the key, these qubits appear maximally mixed.  (Twirling can be thought of as the quantum equivalent of a one-time pad.)
\item Alice transmits the codeword to Bob.  From the secret key, he knows the correct subset of $M$ qubits, and the one-time pad to decode them.
\end{enumerate}
This protocol transmits $(4/3)pN(1-\delta)$ secret qubits from Alice to Bob (Fig. 2). 

If the channel contains intrinsic noise, Alice will first have to encode her $k_s$ stego qubits in an $[[M,k_s]]$ QECC, swap those $M$ qubits for a random subset of $M$ qubits in the codeword, and apply the twirling procedure.  This twirling does not interfere with the error-correcting power of the QECC if Bob knows the key.  Assuming the physical channel is also a DC with error rate $p$, and that Alice emulates a DC with error rate $q$, the effective channel will appear to Eve like a DC with error rate $p + q(1-4p/3) \equiv p + \delta p$.  The rate of transmission $k_s/N$ will depend on the rate of the QECC used to protect the stego qubits.  For a BSC this would be $(1-\delta)(1-h(p))\delta p/(1-2p)$.  However, for most quantum channels (including the DC) the achievable rate is not known.

The secret key is used at two points in these protocols.  First, in step 3 Alice chooses a random subset of $M$ qubits out of the $N$-qubit codeword.  There are $C(N,M)$ subsets, so roughly $\log_2 C(N,M)$ bits are needed to choose one.  Next, in step 4, $2M$ bits of key are used for twirling.  This gives us
\begin{equation}
n_k \approx \log_2 \left(\begin{array}{c} N \\ M \end{array}\right) + 2M
\label{qkeyconsume}
\end{equation}
bits of secret key used.  Define the key consumption rate $\CK=n_k/N$ to be the number of bits of key consumed per qubit that Alice sends through the channel.
We use $M \approx 4qN/3$ and $q \approx \delta p/(1-4p/3)$ to express $\CK$ in terms of $p$, $\delta p$, and $N$ (Fig. 3):
\begin{equation}
\label{KCR}
\CK \approx \log_{2}\left[(4/\beta)^{\beta}(1-\beta N)^{\beta-1}\right]~,\ \ \ \beta \equiv 4\delta{p}/(3-4p)~.
\end{equation} 
Alice can consume fewer bits of key if Bob and she have access to a source that averages to a maximally mixed state.  This would allow them to bypass the twirling procedure.  
The protocols given above perform well in emulating a depolarizing channel.  However, there are far more general channels than these, and the protocols may not work well, or at all, in these cases.  If one has a channel that can be written
\begin{equation}
\rho \rightarrow \CN\rho = (1-p_T + p_E) \CI \rho + p_T \CT\rho + p_E \CE\rho
\label{general_channel}
\end{equation}
where $\CE$ is an arbitrary error operation, one can still use the above protocols to hide approximately $p_T N$ stego bits or qubits, while generating $p_E N$ random errors of type $\CE$.  But for some channels, $p_T$ may be very small or zero.  How should we proceed?  Moreover, hiding stego qubits locally as apparently maximally-mixed qubits sacrifices some potential information.  The {\it location} of the error---that is, the choice of the subset holding the errors---could also be used to convey information, potentially increasing the rate and reducing the amount of secret key or shared entanglement required.

A different approach is instead to encode information in the {\it error syndromes}.  For simplicity, we consider the case when $N$ is large.  In this case, it suffices to consider only {\it typical errors}.  We begin with the case where the physical channel is noise-free.

For large $N$, almost all (probability $1-\epsilon$) combinations of errors on the individual qubits will correspond to one of the set of \textit{typical errors}.  There are roughly $2^{sN}$ of these, and their probabilities $p_e$ are all bounded within a range $2^{-N(s+\delta)} \le p_e \le 2^{-N(s-\delta)}$.  The number $s$ is the entropy of the channel on one qubit; for the BSC $s=h(p)=-p\log_{2}p-(1-p)\log_{2}(1-p)$, and for the DC $s = -(1-p)\log_{2}(1-p)-p\log_{2}p/3$.  We label the typical error operators $E_0, E_1, \ldots, E_{2^{sN}-1}$, and their corresponding probabilities are $p_j$.  A good choice of QECC for the cover text will be able to correct all these errors.  We make the simplifying assumption that the QECC is {\it nondegenerate}, so each typical error $E_j$ has a distinct error syndrome labelled $s_j$.

Ahead of time, Alice and Bob partition the typical errors into $C$ roughly equiprobable sets $S_k$, so that 
\begin{equation}
\sum_{E_j\in S_k} p_j \approx \frac{1}{C},~\forall k~.
\end{equation}
As far as possible, the errors in a given set should be chosen to have roughly equal probabilities.  The maximum of $C$ is roughly $C\approx 2^{N(s-\delta)}$, and $k=0,\ldots,C-1$.  We can now present a new quantum steganographic protocol, using error syndromes to store information.   
\begin{enumerate}
\item Alice prepares $k_c$ qubits of cover text in a state $\ket{\psi_c}$.
\item Alice's secret message is a string of $\log_{2} C \approx N(s-\delta)$ qubits, in a state
\begin{equation}
\ket{\psi_s} = \sum_{k=0}^{C-1} \alpha_k \ket{k}~.
\label{stego_state}
\end{equation}
She ``twirls'' each qubit of this string, using $2N(s-\delta)$ bits of the secret key or shared ebits, to get a maximally mixed state.  To this, she appends $N-k_c-(s-\delta)N$ extra ancilla qubits in the state $\ket0$ to make up a total register of $N-k_c$ qubits.
\item Using the shared secret key, Alice chooses from each set $S_k$ a typical error $E_{j_k}$ with syndrome $s_{j_k}$.  She applies a unitary $U_S$ to the register of $N-k_c$ qubits, that maps $U_S\left(\ket{k}\otimes\ket0^{\otimes N-k_c-sN}\right) = \ket{s_{j_k}}$.  She appends this register to the cover qubits in state $\ket{\psi_c}$, then applies the encoding unitary $U_E$.  Averaging over the secret key, the resulting state will appear to Eve like $\rho \approx \sum_{j=0}^{2^{nS}-1} p_j E_j \ket{\Psi_c}\bra{\Psi_c} E_j^\dagger$, which is effectively indistinguishable from the channel being emulated acting on the encoded cover text.
\item Alice sends this codeword to Bob.  If Eve examines its syndrome, she will find a typical error for the channel being emulated.
\item  Bob applies the decoding unitary $U_D = U_E^\dagger$, and then applies $U_S^\dagger$ (which he knows using the shared secret key).  He discards the cover text and the last $N-k_c-sN$ ancilla qubits, and undoes the twirling operation on the remaining qubits, again using the secret key.  If Eve has not measured the qubits, he will have recovered the state encoded by Alice~\cite{SOM}.
\end{enumerate}
This protocol may easily be used to send classical information by using a single basis state rather than a superposition like (\ref{stego_state}).  The steganographic transmission rate $\CR$ is roughly $\CR\approx s-\delta\rightarrow s$.  The rate of transmission $s$ is higher than the rate $4p/3$ of our first protocol.  This protocol used $2N(s-\delta)$ bits of secret key (or ebits) for twirling in step 2, and roughly $N\delta$ bits of secret key in choosing representative errors $E_{j_k}$ from each set $S_k$ in step 3.  So the key rate is roughly $\CK\approx 2s-\delta\rightarrow 2s$, better than the first protocol in key usage per stego qubit transmitted.  Since almost all the key usage goes to the twirling operation, for sources that are maximally mixed on average the rate of key usage can actually go to zero as $N\rightarrow\infty$.  However, this encoding is much trickier in the case where the channel contains intrinsic noise. 

In principle this quantum steganographic protocol can be used when the channel contains noise.  The steganographic qubits are first encoded in a QECC to protect them against the noise in the channel.  In practice, for many channels this can be difficult:  the effects of errors on the space of syndromes look quite different from a usual additive error channel.  Also, unlike the depolarizing channel, general channels when composed together may change their type.  However, by drawing on codes with suitable properties, the problem of designing steganographic protocols for general channels may be simplified.  We discuss a simple example in the supporting online material (SOM), but the solution for a general channel is a problem for future work.

What is the standard of security for a stego protocol?  There are two obvious considerations.  First, if Eve becomes suspicious, can she read the message?  At the cost of using one-time pads or twirling, Alice and Bob can prevent this from happening.

The more important question is, can Alice and Bob avoid arousing Eve's suspicions in the first place?  To do this, the messages that Alice sends must emulate as closely as possible the channel that Eve expects.  We can make this condition quantitative.  Let $\CE_C$ be the channel on $N$ qubits that Eve expects, and let $\CE_S$ be the effective channel that Alice and Bob produce with their steganographic protocol.  Then the protocol is secure if $\CE_S$ is $\epsilon$-close to $\CE_C$ in the diamond norm $\left\|\CE_S - \CE_C\right\|_\diamond \le \epsilon$ for some small $\epsilon>0$.  The diamond norm is directly related to the probability for Eve to distinguish $\CE_C$ from $\CE_S$ under ideal circumstances (i.e., when she controls both inputs and outputs), and so puts an upper bound on her ability to distinguish them in practice.

For a simple example, the difference between two DCs applied to $N$ qubits has norm
\begin{equation}
\label{dc_diamond}
\left\|\CN_r^{\otimes N} - \CN_{p}^{\otimes N}\right\|_{\diamond} = \sum_{j=0}^N \left(\begin{array}{c} N \\ j \end{array}\right) \left| r^j(1-r)^{N-j} - p^j(1-p)^{N-j}\right| ,
\end{equation}
where $p$ is the error-rate of the channel Eve expects and $r=p+\delta p$ is the error-rate of the steganographic channel that emulates Eve's expected channel.  If we make $\delta p < \epsilon\sqrt{p(1-p)/N}$ then we can make this norm as small as we like, while communicating $O(\delta pN) = O(\epsilon\sqrt{N})$ secret qubits.  This indicates that even if Eve has exact knowledge of the channel, Alice and Bob can in principle send an arbitrarily large (but finite) amount of information without arousing Eve's suspicion, by choosing a sufficiently small $\delta p$ and large $N$~\cite{SOM}.  If Eve's knowledge of the channel is imperfect, Alice and Bob can do even better, communicating steganographic information at a nonzero rate.  If Eve is constantly monitoring the channel over a long period of time, and if she has exact knowledge of the channel then she will eventually learn that Alice and Bob are communicating with each other steganographically.  Moreover, with constant measurement Eve can disrupt the superpositions of the steganographic qubits and prevent any information from ever reaching Bob, effectively flooding the quantum channel with noise.  

If Alice and Bob have shared ebits, they can perform measurements on each of their halves and distill correlated random bits.  Moreover, with shared ebits Alice can send her quantum information to Bob via quantum teleportation by sending only classical bits through the channel.  These classical bits are the result of her measurement on her half of the ebits and her stego qubits.  To Eve who may be monitoring the channel, these bits will look maximally mixed (random).  For her to change the outcome of what Bob receives on his end, Eve would have to disrupt the bits.  So if Eve is measuring the channel continuously, Alice and Bob can still send quantum information to each other using their shared ebits.   


\bibliography{scibib}

\bibliographystyle{Science}

\begin{scilastnote}
\item Acknowledgement.  BAS and TAB would like to thank Daniel Gottesman, Patrick Hayden, Debbie Leung, John Preskill, Stephanie Wehner and Mark Wilde for their useful comments and suggestions at various stages of this work.  This work was funded in part by NSF Grant No.~CCF-0448658.  TAB also acknowledges the hospitality and support of the Kavli Institute for Theoretical Physics.
\end{scilastnote}

\noindent \textbf{Supporting Online Material}\\
SOM Text\\
SOM Figures\\
References


\clearpage

\begin{figure*}[htp]
  \begin{center}
    	 \includegraphics[width = 4.5in]{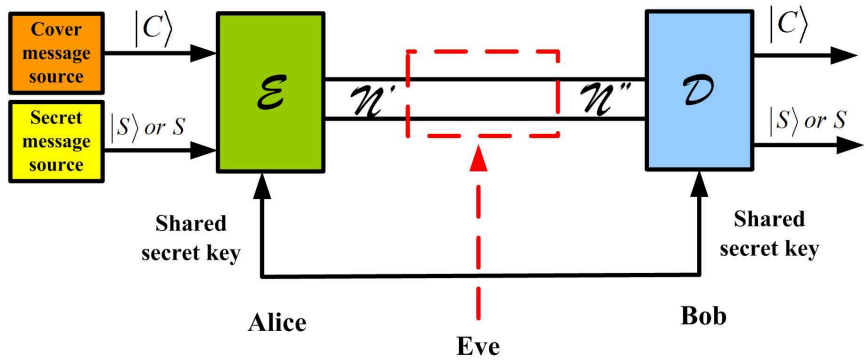}
  \end{center}
  \caption{\small There are three different inputs to the steganographic encoder $\CE$ : a cover-message $\ket{C}$; the secret message that we would like to hide, which can be quantum $\ket{S}$ or classical $S$; a shared secret key which may be quantum (ebit) $\ket{\CK}$ or classical $\CK$.  Eve can monitor some part of the noisy quantum channel $\CN$ shown in the red box.  Bob can decode the steganographic message using the decoder $\CD$ and the shared secret key $\ket{\CK}$ or $\CK$ and recover $\ket{C}$, and $\ket{S}$ or $S$ with very high probability.}
  \label{fig:protocol}
\end{figure*}

\begin{figure*}[htp]
  \begin{center}
    	 \includegraphics[width = 4.5in]{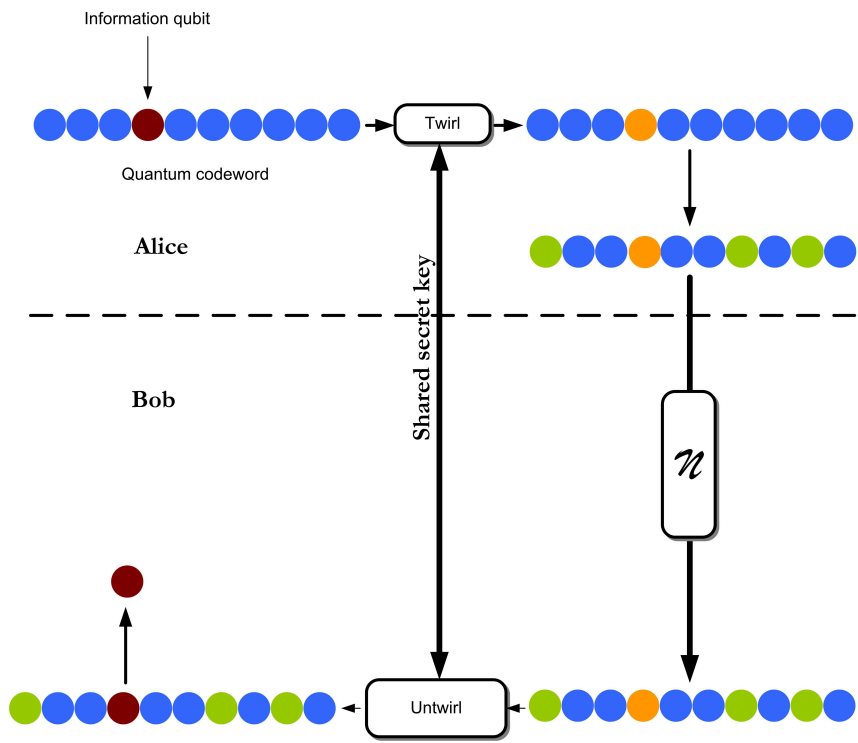}
  \end{center}
  \caption{\small Alice hides her information qubit (solid brown circle) by swapping it in with a qubit of her quantum codeword.  She uses her shared secret key with Bob to determine which qubit to swap.  She uses the shared key again to twirl the information qubit.  She further applies random depolarizing errors to the rest of the qubits of the codeword (shown in green).  She sends the codeword through a depolarizing channel to Bob who uses the shared secret to correctly apply the untwirling operation, followed by locating and swapping out Alice's original information qubit.}
  \label{fig:depolprotocol}
\end{figure*}

\begin{figure*}[htp]
  \begin{center}
    	 \includegraphics[width = 4.5in]{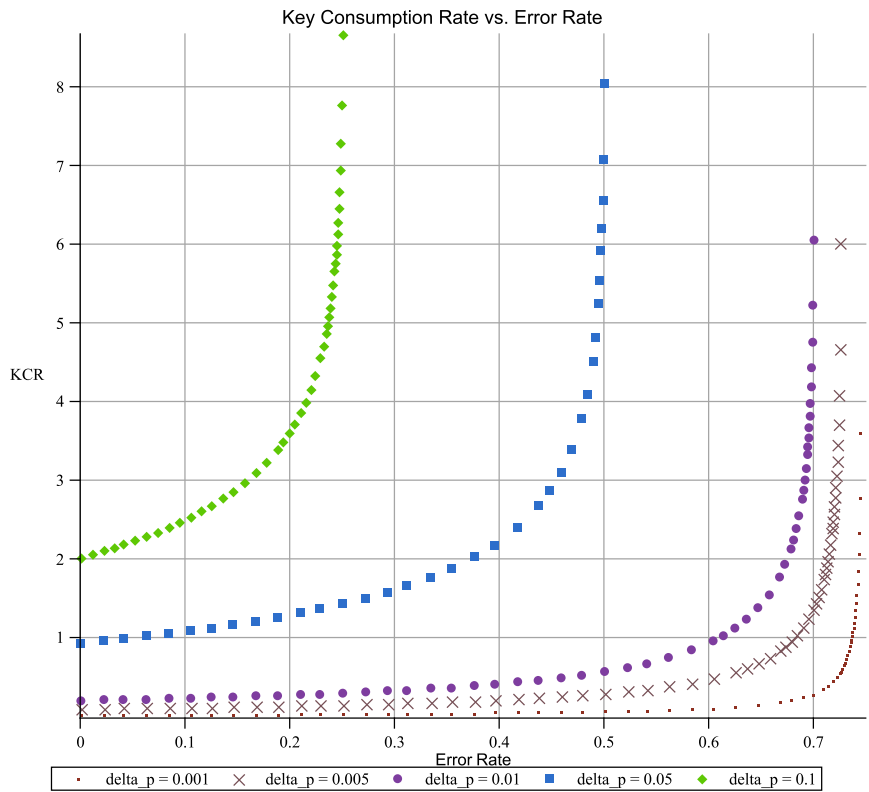}
  \end{center}
  \caption{\small We plot the key consumption rate (KCR) as a function of the error-rate p of the channel.}
  \label{fig:KCR}
\end{figure*}

\clearpage


\section*{Supporting Online Material}
We assume a basic knowledge of quantum information science at the level of~\cite{MikeandIke}.  In this section we gather the definition of the diamond norm and some of its relevant properties to derive the norm of the difference between $N$ uses of two binary-symmetric channels (BSC) and two depolarizing channels (DC).  We refer the reader to John Watrous's lecture notes for the definition and properties of the diamond norm~\cite{Watrous}.  As mentioned in the main text the diamond norm give us a measure of how ``close'' or similar two channels can be when they transform an arbitrary density matrix from one Hilbert space to another.  More formally let $\CN$ be some arbitrary super-operator, and let $\CN:L\left(\CV\right)\rightarrow L\left(\CW \right)$, where $L\left(.\right)$ is a space of linear operators on the Hilbert spaces $\CV$ and $\CW$.  Then one can define the diamond-norm of $\CN$ as:
\begin{equation}
\tag{S1}
\label{eqn:qsecurity-2}
\left\|\CN\right\|_{\diamond} \equiv \left\|I_{L\left(\CV\right)} \otimes \CN\right\|_{tr}~,
\end{equation}
where $\left\|\CN\right\|_{tr}$ is defined as:
\begin{equation}
\tag{S2}
\label{eqn:qsecurity-3}
\left\|\CN\right\|_{tr} \equiv \max\left\{\left\|\CN\left(O\right)\right\|_{tr} : O \in L\left(\CV \right),\left\|O\right\|_{tr}=1 \right\}~.
\end{equation}
The maximization in (S2) is over all density matrices.  When the Hilbert space is infinite dimensional we take the supremum of the set defined in (S2).
\subsection*{Binary Symmetric Channel}
Let $0 < p < 1/2$ be the rate at which Alice flips the qubits of her codeword.  Let $r \equiv p + \delta{p}$ be the rate at which the BSC flips qubits, where $\delta{p}$ is some additional noise which is not under the control of either Alice or Bob.  We assume that $0 < p < r < 1/2$ because at $p = 1/2$ the channel has zero capacity to send information and $p > 1/2$ means that more qubits are being flipped which is unnatural for this channel.  For a single qubit $(N = 1)$ let $\CN_{p}$ be the BSC that Alice applies to an arbitrary single-qubit density operator $\rho$:
\begin{equation}
\tag{S3}
\label{eqn:alicechannel}
\CN_{p}\rho = (1-p)\rho + pX \rho X~,
\end{equation}       
and let $\CN_{r}$ be the actual BSC
\begin{equation}
\tag{S4}
\label{eqn:BSC}
\CN_{r}\rho = (1-r)\rho + rX \rho X~.
\end{equation}       
We can now express the difference of the two channels as:
\begin{equation}
\tag{S5}
\label{eqn:channeldiff}
(\CN_{r} - \CN_{p})\rho = (p-r)\rho + (r-p)X \rho X
\end{equation}
We can express the diamond norm of the difference of the channels $\CN_{p}$ and $\CN_{r}$ as:
\begin{align*}
\label{eqn:diamondnormdiff}
\tag{S6}
 \left\|\CN_{r} - \CN_{p}\right\|_{\diamond} & =  \underset{\rho}{\operatorname{\max}}\left\|(I \otimes (\CN_{r} - \CN_{p}))\rho\right\|_{tr} \\
\tag{S7}
& =  (r - p) \underset{\rho}{\operatorname{\max}}\left\|(I \otimes I)\rho (I \otimes I) - (I \otimes X)\rho(I \otimes X)\right\|_{tr}~. 
\end{align*}
When we substitute $\rho = \psi \otimes \ket{0}\bra{0}$ ($\psi$ is some arbitrary density operator) in the above equation we achieve the maximum.  
\begin{align*}
\label{eqn:singlequbitdiff}
\tag{S8}
\left\|\CN_{r} - \CN_{p}\right\|_{\diamond} & = (r-p)\left\|\psi \otimes \ket{0}\bra{0} - \psi \otimes \ket{1}\bra{1}\right\|_{tr}  \\
\tag{S9}
 & \leq  \left\|\psi \otimes \ket{0}\bra{0}\right\|_{tr} + \left|-1\left\|\right|\psi \otimes \ket{1}\bra{1}\right\|_{tr} \\
\tag{S10} 
 & = (r-p)\left\|\psi\right\|_{tr}\left\|\ket{0}\bra{0}\right\|_{tr} + \left\|\psi\right\|_{tr}\left\|\ket{1}\bra{1}\right\|_{tr} \\
\tag{S11} 
 & = (r-p)(1 + 1) \\
\tag{S12} 
 & =  2(r-p) \\
\tag{S13} 
 & =  2(p + \delta{p} - p) \\
\tag{S14} 
 & =  2\delta{p}.  
\end{align*}
In (S9) we use the triangle inequality and in (S10) we use the fact that for any two linear operators $A$ and $B$, the trace norm of their tensor product is equal to the product of their trace norms, i.e., $\left\|A \otimes B\right\|_{tr} = \left\|A\right\|_{tr}\left\|B\right\|_{tr}$.  We would like an expression for the optimal probability to correctly distinguish two channels. 
\begin{equation}
\label{eqn:p_opt}
\tag{S15}
P_{opt} = \frac{1}{2} + \frac{1}{4}\left\|\CN_{r} - \CN_{p}\right\|_{\diamond}~.
\end{equation}
So for a single-qubit use
\begin{equation}
\label{eqn:p_opt1}
\tag{S16}
P_{opt} = \frac{1}{2}(1+\delta{p})~.
\end{equation}

For the case where we have two qubits, we can write Alice's BSC as:
\begin{equation}
\label{eqn:doubleBSC}
\tag{S17}
(\CN_{p} \otimes \CN_{p})\rho = (1-p)^2\rho + p(1-p)X_{1}\rho X_{1} + p(1-p)X_{2}\rho X_{2} + p^{2}X_{1}X_{2}\rho X_{1}X_{2}~,
\end{equation}
where $X_{1} \equiv X \otimes I$ and $X_{2} \equiv I \otimes X$, and $X_{1}X_{2} \equiv X \otimes X$.  We can similarly calculate $\CN_{1} \otimes \CN_{1}$.  We can now write the difference between the two channels as:
\begin{align*}
\label{eqn:doublediff}
(\CN_{r} \otimes \CN_{r}-\CN_{p} \otimes \CN_{p})\rho & = (r^2-2r+2p-p^2) \\ 
\tag{S18}
& + (r-r^2-p+p^2)(X_{1}\rho X_{1} + X_{2}\rho X_{2}) \\
& + (r^2 - p^2)X_{1}X_{2}\rho X_{1}X_{2}~.
\end{align*}
The diamond norm of the difference between two BSC on two qubits can be expressed as:
\begin{equation}
\label{eqn:twoqubitmax}
\tag{S19}
\left\|\CN_{r} \otimes \CN_{r} - \CN_{p}\otimes \CN_{p}\right\|_{\diamond}  =  \underset{\rho}{\operatorname{\max}}\left\|(I \otimes (\CN_{r} \otimes \CN_{r} - \CN_{p}\otimes \CN_{p}))\rho\right\|_{tr}~. 
\end{equation}  
We use a similar construction from the single-qubit case to maximize the right side of (S17).  Letting $\rho = \psi \otimes \ket{00}\bra{00}$ in (S19), we get:
\begin{equation}
\label{eqn:doublediff1}
\tag{S20}
\left\|\CN_{r} \otimes \CN_{r} - \CN_{p}\otimes \CN_{p}\right\|_{\diamond}  = \left|(1-r)^2-(1-p)^2\right| + 2\left|r(1-r)-p(1-p)\right| + \left|r^2-p^2\right|~.
\end{equation}
Given our constraints that $0 < p < r < 1/2$, the first term on the right side of (S18) is negative while the second and third terms are positive.  This give us:
\begin{align*}
\label{eqn:doublediff2}
\tag{S21}
\left\|\CN_{r} \otimes \CN_{r} - \CN_{p}\otimes \CN_{p}\right\|_{\diamond} & = 2(r-p)(2-r-p) \\
\tag{S22}
& = 2\delta{p}(2-2p-2\delta{p})~.
\end{align*}
So in the double-qubit case $P_{opt}$ is:
\begin{equation}
\label{eqn:p_opt3}
\tag{S23}
P_{opt} = \frac{1}{2}(1 + \delta{p}(2-2p-2r))~.
\end{equation}
If we observe S(20) carefully we find that the terms are distributed binomially.  For the case where we have $N$ qubits, we can use $\rho = \psi \otimes \ket{00\cdots 0}\bra{00\cdots 0}$ to maximize the diamond norm for $N$ uses of BSC to get:
\begin{equation}
\label{eqn:diamond_N}
\tag{S24}
\left\|\CN_r^{\otimes N} - \CN_{p}^{\otimes N}\right\|_{\diamond} = \sum_{j=0}^N \left(\begin{array}{c} N \\ j \end{array}\right) \left| r^j(1-r)^{N-j} - p^j(1-p)^{N-j}\right|~.
\end{equation} 

\subsection*{Depolarizing Channel}
The calculation of the diamond norm of the difference between $N$ uses of two depolarizing channels (DC) is similar to the calculation of BSC that we performed in the previous section.  The expression for the channel is
\begin{equation}
\label{eqn:alicedepol}
\tag{S25}
\CN_{p}\rho = (1-p)\rho + (p/3_(X\rho X + Y\rho Y + Z\rho Z)~.
\end{equation}
Eve sees a channel with a somewhat higher rate $r=p+\delta p$.  As in the BSC case we assume that $0 < p < r < 1/2$.  For $N = 2$ case the difference between the two depolarizing channels is:
\begin{align*}
\label{eqn:depoldiff1}
(\CN_{r}\otimes\CN_{r} - \CN_{p}\otimes\CN_{p})\rho & = ((1-r)^2-(1-p)^2)\rho \\ 
\tag{S27}
& + ((1-r)(r/3) - (1-p)(p/3))(X_{1}\rho X_{1} + \cdots + Z_{2}\rho Z_{2}) \\
& + ((r/3)^2 - (p/3)^2)(X_{1}X_{2}\rho X_{1}X_{2} + \cdots + Z_{1}Z_{2}\rho Z_{1}Z_{2})~.
\end{align*}
As in the BSC case we can express the diamond norm as in (S19).  The density matrix that maximizes the trace norm is $\rho = \psi \otimes \ket{\Phi^{+}}\bra{\Phi^{+}}$, where $\ket{\Phi^{+}} = 1/\sqrt{2}(\ket{00} + \ket{11})$, and $\psi$ is some arbitrary single-qubit density operator. 
\begin{align*}
\left\|\CN_{r} \otimes \CN_{r} - \CN_{p}\otimes \CN_{p}\right\|_{\diamond} & = \left|(1-r)^2 - (1-p)^2\right| \\ 
\tag{S28}
& + 6\left|(1-r)(r/3) - (1-p)(p/3)\right| \\
& + 9\left|(r/3)^2 - (p/3)^2\right| \\
& = \left|(1-r)^2-(1-p)^2\right| + 2\left|(1-r)r - (1-p)p\right| + \left|r^2 - p^2\right|~.
\end{align*}
After evaluating the absolute value terms, we get:
\begin{align*}
\label{eqn:depol2}
\left\|\CN_{r} \otimes \CN_{r} - \CN_{p}\otimes \CN_{p}\right\|_{\diamond} & = 2(r-p)(2-r-p) \\
\tag{S29}
& = 2\delta{p}\bigg(2 - 2p - \delta{p}\bigg)~.
\end{align*}
So,
\begin{equation}
\label{eqn:depol3}
\tag{S30}
P_{opt} = \frac{1}{2} + \frac{1}{2}\delta{p}\bigg(2 - 2p - \delta{p}\bigg)~.
\end{equation}
For the general case for $N$ uses of the depolarizing channel we may write the diamond norm as:
\begin{equation}
\label{eqn:depoldiamond_N}
\tag{S31}
\left\|\CN_r^{\otimes N} - \CN_{p}^{\otimes N}\right\|_{\diamond} = \sum_{j=0}^N \left(\begin{array}{c} N \\ j \end{array}\right) \left| r^j(1-r)^{N-j} - p^j(1-p)^{N-j}\right|~,
\end{equation}
which is exactly the same expression as for the BSC.

\subsection*{Achievable Rate for Protocol 2}
We will work out the simplest example---the BSC in the case where the physical channel is noise-free.  The errors in the codewords that Alice sends to Bob are binomially distributed.  Let $pN$ be the mean of this distribution and let the variance be $pN\delta$, where $0 < \delta \ll 1$.  Here $N$ is the length of each of codeword.  Let 
\begin{equation}
\label{eqn:rate1}
\tag{S32}
p_{k} = {N \choose{k}}p^{k}(1-p)^{N-k}
\end{equation}
be the errors that Alice applies to her codewords.  For each $k$ from $Np(1-\delta)$ to $Np(1+\delta)$ choose $C_{k}$ strings of weight $k$.  Let 
\begin{equation}
\label{eqn:rate2}
\tag{S33}
C = \sum_{k = Np(1-\delta)}^{Np(1+\delta)} C_{k}~.
\end{equation}    
Let these sets of strings be called $S_{k}$, and 
\begin{equation}
\label{eqn:rate3}
\tag{S34}
S = \cup_{k} S_{k}
\end{equation}
So the total number of strings in the set $S$ is $C$.  Define the probability $q \equiv 1/C$.  Then we want to satisfy $qC_{k} = C_{k}/C = p_{k}$.  Clearly we must have $C_{k} \leq N \choose{k}$, for all $k$.  This implies that:
\begin{align*}
\label{eqn:rate4}
C_{k}p^{k}(1-p)^{N-k} & \leq {N \choose{k}}p^{k}(1-p)^{N-k}  \\
\Rightarrow C_{k}p^{k}(1-p)^{N-k} & \leq C_{k}q \\
\Rightarrow p^{k}(1-p)^{N-k} & \leq q
\end{align*}
We want $C$ to be as large as possible, which means we want $q$ to be as small as possible.  This constraint then gives us
\begin{align*}
q & = p^{Np(1-\delta)}(1-p)^{N(1-p+p\delta)} \\
\Rightarrow C & = 1/q \\
\Rightarrow C & = p^{-Np(1-\delta)}(1-p)^{-N\left(1-p+p\delta\right)} 
\end{align*}
The number of bits that Alice can send is, therefore
\begin{align*}
M & = \log_{2}C \\
& = N(-p\log_{2}p-(1-p)\log_{2}(1-p)+\delta(p\log_{2}p-p\log_{2}(1-p))) \\
\tag{S35}
& = N(h(p)-p\delta\log_{2}((1-p)/p)) 
\end{align*}
So with this encoding Alice can send almost $Nh(p)$ bits.

\subsection*{Diamond norm for protocol 2}
Again we consider the simplest case of the BSC.  Let $N$ be sufficiently large so that the total probability of the typical errors is $> 1-\epsilon$, and these typical errors have weight $k$ in the range $Np(1-\delta) \le k \le Np(1+\delta)$.  We divide up all errors of weight $k$ into $C_k$ partitions containing
\[
n_k\approx \frac{{N\choose{k}}}{C_k}\approx \left(\frac{1-p}{p}\right)^{k-Np(1-\delta)}
\]
errors each.  Within each set the errors are all equally likely to be chosen.  However, because the number of errors is unlikely to divide exactly evenly into $C_k$ sets, the probabilities $q_k$ of an error of weight $k$ will be slightly different from the probability $p_k = p^k(1-p)^{N-k}$ of the binomial distribution.  We can put a (not-very-tight) bound on this difference:
\begin{equation}
\tag{S36}
|q_k - p_k| < \frac{p_k}{\left(\frac{1-p}{p}\right)^{k-Np(1-\delta)}-1}
< \frac{1-p}{1-2p} p^{2k}(1-p)^{N-2k} .
\end{equation}
Plugging this into the expression for the diamond norm, we get
\begin{align*}
||\CN_p^{\otimes N} - \CN_{\rm enc}||_\diamond < & \epsilon + \sum_{k=Np(1-\delta)+1}^{Np(1+\delta)} {N\choose{k}} |p_k-q_k| \\ \tag{S37}
< & \epsilon + \left(\frac{1-p}{1-2p}\right) \left(\frac{p}{1-p}\right)^{Np(1-\delta)}
\left(\frac{1-2p+2p^2}{1-p}\right)^N ~,
\end{align*}
which is exponentially small in $N$.

\subsection*{Error-correction for protocol 2 with a noisy channel}
Since errors can act in a complicated manner on the space of syndromes, it is not entirely clear what the optimal encoding is even for a simple channel.  Here we present one encoding for the BSC that gives an achievable rate in the limit of large $N$, but it is quite likely that higher rates are possible.

In the noiseless case, it is possible to use the $C(N,M)$ strings of weight $M$ as a code---each string represents one possible weight-$M$ error.  If we then apply a BSC with probability $p$, on average $Np$ bits would be flipped.  If $Np \ll M$ then one can keep only a subset of the weight-$M$ strings, separated by a distance $>2Np$.

This encoding quickly becomes inefficient as $p$ gets larger.  Using the shared secret key, Alice can instead chose only a subset of the $N$ bits  to hold the codewords.  If this subset includes $N'$ bits, then the errors on the remaining $N-N'$ bits are irrelevant and do not need to be corrected.  The limit of this would be similar to encoding 1 in the paper, where $N' \approx 2M$.

Let $N'=qN$ for some $0<q\le1$.  The number of strings of weight $M$ is $C(qN,M)$, and there will be an average number of bit flips $pqN$ on the relevant portion of the codeword.  Keep a subset of these codewords separated by distance $2pqN$.  Decoding is done by finding the closest codeword to the output string.

As $N,M\rightarrow\infty$ then the number of codewords will go like
\[
C(N,M,p,q) \sim \frac{{qN\choose{M}}}{{qN\choose{pqN}}}~.
\]
The number of bits will be $\log_2 C(N,p,q)$.

Since $q$ is a parameter we can choose freely, we choose it to maximize the rate $\CR(N,M,p,q) \equiv (1/N) \log_2 C(N,M,p,q)$.  Using the Stirling approximation, differentiating with respect to $q$, and setting the result equal to 0, we can solve for $q$:
\[
q = \frac{M}{N}\left(\frac{2^{h(p)}}{2^{h(p)}-1}\right) ~.
\]
We can then plug this back into the formula for $\CR$.  If the physical channel has error rate $p$ and Alice is attempting to emulate a channel with error rate $p+\delta p$, then $M=N\delta p/(1-2p)$.  This gives us the following expression for the rate:
\[
\CR(p,\delta p) = - \frac{\delta p}{1-2p} \log_2\left(2^{h(p)}-1\right) .
\]
We can compare this to the rate from encoding 1, which for the BSC is $2\delta p(1-h(p))/(1-2p)$.  It is not hard to see that $\CR(p,\delta p)$ above approaches this rate as $p\rightarrow1/2$ (and both rates go to zero), but as $p\rightarrow0$ this encoding does considerably better than encoding 1.  It is quite likely, however, that there may be even more efficient encodings.

\subsection*{Shared classical secret key vs. shared ebits}

If Alice and Bob share a secret, random key, they can use the steganographic encodings described in the paper.  Shared entanglement (ebits) can act as a resource in the same way---by measuring the two halves of a maximally entangled pair of qubits $(|00\rangle + |11\rangle)/\sqrt2$ Alice and Bob can generate a shared secret bit.

However, the use of ebits does open up an additional possibility beyond what can be done with a classical key.  Instead of sending quantum information through the channel, Alice can instead {\it teleport} qubits to Bob.  Teleportation consumes one ebit and requires the transmission of two classical bits for each qubit teleported.  These classical bits can be sent through the channel steganographically.  Because these bits are perfectly random, no one-time pad or twirling is needed.  And because they are purely classical information, they are not disrupted if Eve chooses to measure the error syndromes, as a general quantum state would be.  In this sense, quantum steganography with shared ebits is more powerful than quantum steganography with a shared classical key.

\bibliography{scibib}

\bibliographystyle{Science}

\end{document}